\documentclass[aps,preprint]{revtex4}%
\usepackage{amsfonts}
\usepackage{amsmath}
\usepackage{amssymb}
\usepackage{graphicx}%
\begin{document}
\title{Ground state energy of spin-1/2 fermions in the unitary limit}
\author{Dean Lee}
\affiliation{Department of Physics, North Carolina State University, Raleigh, NC 27695}

\begin{abstract}
We present lattice results for the ground state energy of a spin-1/2 fermion
system in the unitary limit, where the effective range of the interaction is
zero and the scattering length is infinite. \ We compute the ground state
energy for a system of 6, 10, 14, 18, and 22 particles, with equal numbers of
up and down spins in a periodic cube.\ \ We estimate that in the limit of
large number of particles, the ground state energy is $0.25(3)$ times the
ground state energy of the free Fermi system.

\end{abstract}
\maketitle

\section{Introduction}

The unitary limit of spin-1/2 or two-component fermions has attracted much
recent experimental and theoretical interest from several branches of physics.
\ The term unitary limit or unitary regime refers to the idealized limit where
the effective range of the interaction is zero and the scattering length is
infinite. \ Much of the interest has been spurred by experimental advances in
the trapping of ultracold atomic Fermi gases. \ Starting with a dilute Fermi
gas, where the effective range of the interaction is negligible compared with
the interparticle spacing, one can reach the unitary limit by tuning the
scattering length to infinity using a Feshbach resonance
\cite{O'Hara:2002,Gupta:2002,Regal:2003,Bourdel:2003,Gehm:2003,Bartenstein:2004,Kinast:2005}%
.

In nuclear physics the unitary limit is directly relevant to the properties of
cold dilute neutron matter. \ The neutron scattering length is roughly $-18$
fm while the effective range is $2.8$ fm. \ Therefore the unitary limit is
approximately realized when the interparticle spacing is about $5-10$ fm,
roughly $0.5\%-5\%$ of normal nuclear matter density, which is believed to be
relevant to the physics of the inner crust of neutron stars
\cite{Pethick:1995di}.

At zero temperature in the unitary limit there are no dimensionful parameters
other than the particle density. \ Therefore the ground state energy of the
system should obey a simple relation $E^{0}=\xi E^{0,\text{free}}$, where
$E^{0,\text{free}}$ is the ground state energy of a non-interacting Fermi gas
and $\xi$ is a dimensionless constant. \ Recent experiments have measured the
expansion of $^{6}$Li in the unitary limit released from a harmonic trap
\cite{Bartenstein:2004,Kinast:2005}. \ Based on a Thomas-Fermi model, the
measured values for $\xi$ are $0.51(4)$ \cite{Kinast:2005} and $0.32_{-13}%
^{+10}$ \cite{Bartenstein:2004}. \ The discrepancy between these two recent
measurements and with larger values for $\xi$ reported in earlier experiments
\cite{O'Hara:2002,Bourdel:2003,Gehm:2003} suggest further experimental work is
needed, as well as a better theoretical understanding of Thomas-Fermi theory
and other local density approximations in the unitary limit.

There have been several recent analytic calculations of $\xi$ using various
techniques such as BCS saddle point and variational approximations, Pad\'{e}
approximations, mean field theory with pairing, and dimensional expansions
\cite{Engelbrecht:1997,Baker:1999dg,Heiselberg:1999,Perali:2004,Schafer:2005kg}%
. \ The values for $\xi$ vary from roughly $0.3$ to $0.6$. \ Fixed-node
Green's function Monte Carlo calculations have found $\xi$ to be $0.44(1)$
\cite{Carlson:2003z} and $0.42(1)$ \cite{Astrakharchik:2004}. \ A recent
estimate based on Kohn-Sham theory for the two-fermion system in a harmonic
trap yields a value of $0.42$ \cite{Papenbrock:2005}. \ Recently there have
also been simulations of spin-1/2 fermions on the lattice in the unitary limit
\cite{Wingate:2005xy,Bulgac:2005a,Lee:2005is,Lee:2005it}. \ While the
simulations are at nonzero temperature, the results can be extrapolated to
provide an estimate for $\xi$. \ The results of \cite{Bulgac:2005a} produced a
value for $\xi$ similar to the fixed-node results, while \cite{Lee:2005it}
established a bound, $0.07\leq\xi\leq0.42$.

In this paper we describe a new calculation of $\xi$ on the lattice at zero
temperature and fixed particle number. \ In contrast with Green's function
Monte Carlo, the method we describe is free from the fermion sign problem.
\ Because of this we can eliminate systematic errors due to fermion nodal
constraints. \ Using a new algorithm which combines endpoint correlation
function importance sampling and non-local Monte Carlo updating, we measure
the ground state energy for 6, 10, 14, 18, and 22 particles, with equal
numbers of up and down spins in a periodic cube.\ \ We estimate that in the
limit of large number of particles, the ground state energy is $0.25(3)$ times
the ground state energy of the free Fermi system.

\section{Lattice formalism}

We refer to the state with $N$ spin-up fermions and $N$ spin-down fermions as
an $N,N$ state. \ We use the same lattice action as described in
\cite{Lee:2005is,Lee:2005it,Lee:2004qd}. \ Since we will be working at fixed
particle number we can set the chemical potential to zero. \ Throughout we use
dimensionless parameters and operators, which correspond with physical values
multiplied by the appropriate power of the spatial lattice spacing $a$. \ For
quantities in physical units we use the superscript `phys'. \ The lattice
action has the form%
\begin{align}
&  \sum_{\vec{n},i}\left[  c_{i}^{\ast}(\vec{n})c_{i}(\vec{n}+\hat
{0})-e^{\sqrt{-C\alpha_{t}}s(\vec{n})+\frac{C\alpha_{t}}{2}}(1-6h)c_{i}^{\ast
}(\vec{n})c_{i}(\vec{n})\right] \nonumber\\
&  -h\sum_{\vec{n},l_{s},i}\left[  c_{i}^{\ast}(\vec{n})c_{i}(\vec{n}+\hat
{l}_{s})+c_{i}^{\ast}(\vec{n})c_{i}(\vec{n}-\hat{l}_{s})\right]  +\frac{1}%
{2}\sum_{\vec{n}}\left(  s(\vec{n})\right)  ^{2}. \label{standard}%
\end{align}
Here $\vec{n}$ labels the sites of a $3+1$ dimensional lattice, $\hat{l}%
_{s}\ (s=1,2,3)$ are the spatial lattice unit vectors, $\hat{0}$ is a temporal
lattice unit vector, and $i$ labels the two spin components of the fermion.
\ The temporal lattice spacing is $a_{t}$, and $\alpha_{t}=a_{t}/a$ is the
ratio of the temporal to spatial lattice spacing. \ We have also defined
$h=\alpha_{t}/(2m)$, where $m$ is the fermion mass in lattice units.

Our choices for the physical values of the fermion mass and lattice spacings
are irrelevant to the universal physics of the unitary limit. \ Nevertheless
it is convenient to assign some concrete values to these parameters. \ The
values we choose are motived by the dilute neutron system. We use a fermion
mass of $939$ MeV and lattice spacings $a=(50$ MeV$)^{-1}$, $a_{t}=(24$
MeV$)^{-1}$. \ Our spatial geometry is a periodic cube of length $L$ lattice
units. \ The Grassmann fields are denoted by $c_{i}(\vec{n})$. $\ s(\vec{n})$
is an auxiliary Hubbard-Stratonovich field which upon integration reproduces
the attractive contact interaction between up and down spins. \ The
interaction coefficient for these lattice spacings is computed by summing
two-particle scattering bubble diagrams. \ The details of the calculation are
given in \cite{Lee:2004qd}. \ Taking the scattering length $a_{\text{scatt}%
}\rightarrow\infty$, we find in the unitary limit, $C^{\text{phys}%
}=-1.203\times10^{-4}$ MeV$^{-2}.$

In order to compute the ground state energy, we consider the correlation
function%
\begin{equation}
Z_{N,N}(t)=\left\langle \Psi_{N,N}^{0}\right\vert e^{-Ht}\left\vert \Psi
_{N,N}^{0}\right\rangle ,
\end{equation}
where the initial/final state $\left\vert \Psi_{N,N}^{0}\right\rangle $ is the
normalized Slater determinant ground state for the free $N,N$ particle system.
\ We refer to $t$ as the Euclidean time. \ We define%
\begin{equation}
E_{N,N}(t)=-\frac{\partial}{\partial t}\left[  \ln Z_{N,N}(t)\right]  .
\end{equation}
Then as $t\rightarrow+\infty$, $E_{N,N}(t)$ converges to $E_{N,N}^{0}$, the
ground state energy of the interacting $N,N$ particle system. \ The only
assumption is that the ground state has a nonvanishing overlap with the ground
state of the non-interacting system.

For the non-interacting system we find $E_{N,N}^{\text{free}}(t)=E_{N,N}%
^{0,\text{free}}$, where $E_{N,N}^{0,\text{free}}$ is the free Fermi ground
state energy. \ It is useful to define the dimensionless function%
\begin{equation}
\xi_{N,N}(t)=\frac{E_{N,N}(t)}{E_{N,N}^{0,\text{free}}}.
\end{equation}
In the unitary limit $\xi_{N,N}(t)$ depends only on the dimensionless
combination $\frac{t}{mL^{2}}$. \ We can define the Fermi energy,%
\begin{equation}
E_{F}=\frac{k_{F}^{2}}{2m}=\frac{1}{2m}\left(  3\pi^{2}\rho\right)
^{2/3}\simeq7.596\frac{N^{2/3}}{mL^{2}},
\end{equation}
where $k_{F}$ is the Fermi momentum. \ In the unitary limit we can regard
$\xi_{N,N}(t)$ as a function of $E_{F}t$, and for sufficiently large $N$ it
should approach a common asymptotic form.

The conversion of the Grassmann lattice action to a worldline formalism at
fixed particle number has been detailed in \cite{Borasoy:2005yc}. \ We use the
same transfer matrix derived there, except in this case we keep the auxiliary
Hubbard-Stratonovich field and calculate the integral over auxiliary field
configurations,%
\begin{equation}
Z_{N,N}(t)\varpropto\int Ds\;e^{-\frac{1}{2}\sum_{\vec{n}}\left(  s(\vec
{n})\right)  ^{2}}\left\langle \Psi_{N,N}^{0}\right\vert Te^{-H(s)t}\left\vert
\Psi_{N,N}^{0}\right\rangle .
\end{equation}
The advantage of the auxiliary field formalism is that $H(s)$ consists of only
single-body operators interacting with the background auxiliary field. \ We
can therefore compute the full $N,N$-body matrix element as the square of the
determinant of the single-particle matrix elements,%
\begin{align}
\left\langle \Psi_{N,N}^{0}\right\vert Te^{-H(s)t}\left\vert \Psi_{N,N}%
^{0}\right\rangle  &  \propto\left[  \det M(s,t)\right]  ^{2},\\
M_{ij}(s,t) &  =\left\langle p_{i}\right\vert Te^{-H(s)t}\left\vert
p_{j}\right\rangle ,
\end{align}
where $i,j$ go from $1$ to $N.$ \ The states $\left\vert p_{j}\right\rangle $
are single-particle momentum states comprising our Slater determinant
initial/final state. \ The square of the determinant arises from the fact that
the single-body operators in $H(s)$ are the same for both up and down spins.
\ Since the square of the determinant is nonnegative, there is no sign problem
in this formalism.

Our notation, $Te^{-H(s)t}$, is shorthand for the time-ordered product of
single-body transfer matrices at each time step,%

\begin{equation}
Te^{-H(s)t}=M(L_{t}-1)\cdot\ldots\cdot M(n_{t})\cdot\ldots\cdot M(1)\cdot
M(0),
\end{equation}
where $L_{t}$ is the total number of lattice time steps and $t=L_{t}a_{t}$.
\ If the particle stays at the same spatial lattice site from time step
$n_{t}$ to $n_{t}+1$, then the corresponding matrix element of $M(n_{t})$ is%
\begin{equation}
e^{\sqrt{-C\alpha_{t}}s(\vec{n})+\frac{C\alpha_{t}}{2}}(1-6h).
\end{equation}
If the particle hops to a neighboring lattice site from time step $n_{t}$ to
$n_{t}+1$ then the corresponding matrix element of $M(n_{t})$ is $h$. \ All
other elements of $M(n_{t})$ are zero.

\section{Endpoint importance sampling and hybrid Monte Carlo}

While there is no sign problem, the calculation of $Z_{N,N}(t)$ is not
straightforward. \ Due to the large coupling strength in the unitary limit,
fluctuations in $\left[  \det M(s,t)\right]  ^{2}$ are very large. \ We deal
with this problem by sampling configurations according to the weight%
\begin{equation}
\exp\left\{  -\frac{1}{2}\sum_{\vec{n}}\left(  s(\vec{n})\right)  ^{2}%
+2\log\left[  \left\vert \det M(s,t_{\text{end}})\right\vert \right]
\right\}  ,
\end{equation}
where $t_{\text{end}}$ is the largest Euclidean time at which we wish to
measure $Z_{N,N}(t)$. \ It is most efficient to sample configurations using a
non-local updating method called hybrid Monte Carlo \cite{Duane:1987de}. \ A
description of the hybrid Monte Carlo method as applied to the grand canonical
ensemble of spin-1/2 neutrons can be found in \cite{Lee:2004qd}. \ For this
fixed particle number simulation we compute molecular dynamics trajectories
for%
\begin{equation}
H(s,p)=\frac{1}{2}\sum_{\vec{n}}\left(  p(\vec{n})\right)  ^{2}+V(s),
\end{equation}
where $p(\vec{n})$ is the conjugate momentum for $s(\vec{n})$ and%
\begin{equation}
V(s)=\frac{1}{2}\sum_{\vec{n}}\left(  s(\vec{n})\right)  ^{2}-2\log\left[
\left\vert \det M(s,t_{\text{end}})\right\vert \right]  .
\end{equation}
The steps of the algorithm are as follows.

\begin{itemize}
\item[Step 1:] Select an arbitrary initial configuration $s^{0}(\vec{n})$.

\item[Step 2:] Select $p^{0}(\vec{n})$ according to the Gaussian random
distribution%
\begin{equation}
P(p^{0}(\vec{n}))\propto\exp\left[  -\frac{1}{2}(p^{0}(\vec{n}))^{2}\right]  .
\end{equation}

\item[Step 3:] Let%
\begin{equation}
\tilde{p}^{0}(\vec{n})=p^{0}(\vec{n})-\frac{\varepsilon}{2}\left[
\frac{\partial V(s)}{\partial s(\vec{n})}\right]  _{s=s^{0}}%
\end{equation}
for some small positive $\varepsilon$. \ In computing the derivative of $V$,
we use the fact that%
\begin{align}
\frac{\partial V(s)}{\partial s(\vec{n})}  & =s(\vec{n})-\frac{2}{\det M}%
\sum_{k,l}\frac{\partial\det M}{\partial M_{kl}}\frac{\partial M_{kl}%
}{\partial s(\vec{n})}\nonumber\\
& =s(\vec{n})-2\sum_{k,l}\left[  M^{-1}\right]  _{lk}\frac{\partial M_{kl}%
}{\partial s(\vec{n})}.
\end{align}

\item[Step 4:] For $j=0,1,...,N-1$, let%
\begin{align}
s^{j+1}(\vec{n}) &  =s^{j}(\vec{n})+\varepsilon\tilde{p}^{j}(\vec{n}),\\
\tilde{p}^{j+1}(\vec{n}) &  =\tilde{p}^{j}(\vec{n})-\varepsilon\left[
\frac{\partial V(s)}{\partial s(\vec{n})}\right]  _{s=s^{j+1}}.
\end{align}

\item[Step 5:] Let%
\begin{equation}
p^{N}(\vec{n})=\tilde{p}^{N}(\vec{n})+\frac{\varepsilon}{2}\left[
\frac{\partial V(s)}{\partial s(\vec{n})}\right]  _{s=s^{N}}.
\end{equation}

\item[Step 6:] Select a random number $r\in$ $[0,1).$ \ If
\begin{equation}
r<\exp\left[  -H(s^{N},p^{N})+H(s^{0},p^{0})\right]
\end{equation}
then let%
\begin{equation}
s^{0}(\vec{n})=s^{N}(\vec{n}).
\end{equation}
Otherwise leave $s^{0}$ as is. \ In either case go back to Step 2.
\end{itemize}

For each configuration the observables that we calculate are%
\begin{equation}
O(s,t)=\frac{\left[  \det M(s,t)\right]  ^{2}}{\left[  \det M(s,t_{\text{end}%
})\right]  ^{2}},
\end{equation}
for $t<t_{\text{end}}$. \ By taking the ensemble average of $O(s,t)$ we are
able to calculate%
\begin{equation}
\frac{Z_{N,N}(t)}{Z_{N,N}(t_{\text{end}})}.
\end{equation}
Fluctuations in the observable $O(s,t)$ are manageable in size so long as
$t_{end}-t$, the temporal separation from the endpoint, is not too large. This
is typically not a problem since only a small window in $t$ near $t_{end}$ is
needed to calculate $-\frac{\partial}{\partial t}\left[  \ln Z_{N,N}%
(t)\right]  $ at $t=t_{\text{end}}$. \ For each simulation we compute roughly
$2\times10^{5}$ hybrid Monte Carlo trajectories, split across four processors
running completely independent trajectories. \ Averages and errors are
computed by comparing the results of each processor.

\section{Error estimates}

We use double precision arithmetic to compute $\det M(s,t_{\text{end}})$ and
$O(s,t)$. \ We carefully monitor any systematic errors produced by double
precision round off error and exceptional configurations. \ To do this we
introduce a small positive parameter $\epsilon$ and reject any hybrid Monte
Carlo trajectories which generate a configuration with%
\begin{equation}
\left\vert \det M\right\vert <\epsilon^{N}\prod\limits_{i=1,...,N}\left\vert
M_{ii}\right\vert \text{.}%
\end{equation}
We then take the limit $\epsilon\rightarrow0^{+}$ to determine if poorly
condition matrices make any detectable contribution to our observable. \ We
consider values for $\epsilon$ as small as $10^{-6}$. \ If as we take
$\epsilon\rightarrow0^{+}$ any systematic error can be detected above the
stochastic error level, then we throw out the measurement and do not include
it in the final results. \ The error bars we present are therefore estimates
of the total error for each lattice system. \ There are no additional errors
other than the lattice spacing dependence.

Errors due to finite lattice spacing can be estimated by the spread of
$\xi_{N,N}(t)$ at fixed $N$ for different lattice lengths $L$. \ By fitting
the data for different $L$ to a single function $\xi_{N,N}(t)$, we can compute
the uncertainty in the fit parameters and estimate the error of the
measurement in the continuum limit. \ Therefore the error bars on our final
results include both systematic and stochastic errors for each lattice system
as well as systematic errors due to lattice spacing dependence.

\section{Results}

As a first test we consider the $1,1$ particle system. \ In this case the free
Fermi ground state energy $E_{1,1}^{0,\text{free}}$ is zero. \ So instead of
dividing by $E_{1,1}^{0,\text{free}}$, we measure the dimensionless quantity
$mL^{2}E_{1,1}(t)$. \ In Table 1 $mL^{2}E_{1,1}(t)$ is shown for various $t$
and $L$. \ We also show the results for the dimensionless interacting ground
state energy, $mL^{2}E_{1,1}^{0}$, as computed by summing lattice bubble
diagrams.%
\[%
\genfrac{}{}{0pt}{}{\text{Table 1: \ }mL^{2}E_{1,1}(t)\text{ for various
}t\text{ and }L}{%
\begin{tabular}
[c]{|l|l|l|l|l|l|l|}\hline
$L$ & $12a_{t}$ & $24a_{t}$ & $36a_{t}$ & $48a_{t}$ & $60a_{t}$ &
$mL^{2}E_{1,1}^{0}$\\\hline
$4$ & $-2.7(1)$ & $-3.4(1)$ & $-3.4(1)$ & $-3.4(1)$ & $-3.5(1)$ &
$-3.57$\\\hline
$5$ & $-2.5(1)$ & $-3.1(1)$ & $-3.5(1)$ & $-3.5(1)$ & $-3.5(1)$ &
$-3.60$\\\hline
$6$ & $-2.0(1)$ & $-2.7(1)$ & $-3.1(1)$ & $-3.6(1)$ & $-3.5(1)$ &
$-3.62$\\\hline
\end{tabular}
\ \ \ }%
\]
\ For large $t$ we see that $mL^{2}E_{1,1}(t)$ matches $mL^{2}E_{1,1}^{0}$
within error bars. \ For large $L$ the lattice discretization error should be
negligible and $mL^{2}E_{1,1}^{0}$ should approach the unitary limit result,%
\begin{equation}
4\pi^{2}d_{1}\simeq4\pi^{2}(-0.095901)\simeq-3.7860,
\end{equation}
where $d_{1}$ is the large scattering length expansion coefficient in a
periodic cube as defined in \cite{Beane:2003da}.\ \ We see that $mL^{2}%
E_{1,1}^{0}$ is within a few percent of the unitary limit value for $L=4,5,6$.

\begin{figure}
[ptb]
\begin{center}
\includegraphics[
height=2.6238in,
width=3.6677in
]%
{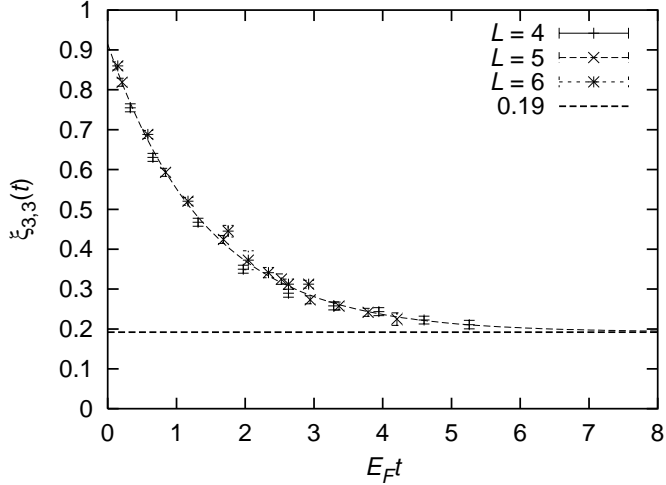}%
\caption{$\xi_{3,3}(t)$ versus $E_{F}t$ for $L=4,5,6$.}%
\label{3+3}%
\end{center}
\end{figure}

\begin{figure}
[ptbptb]
\begin{center}
\includegraphics[
height=2.6238in,
width=3.6677in
]%
{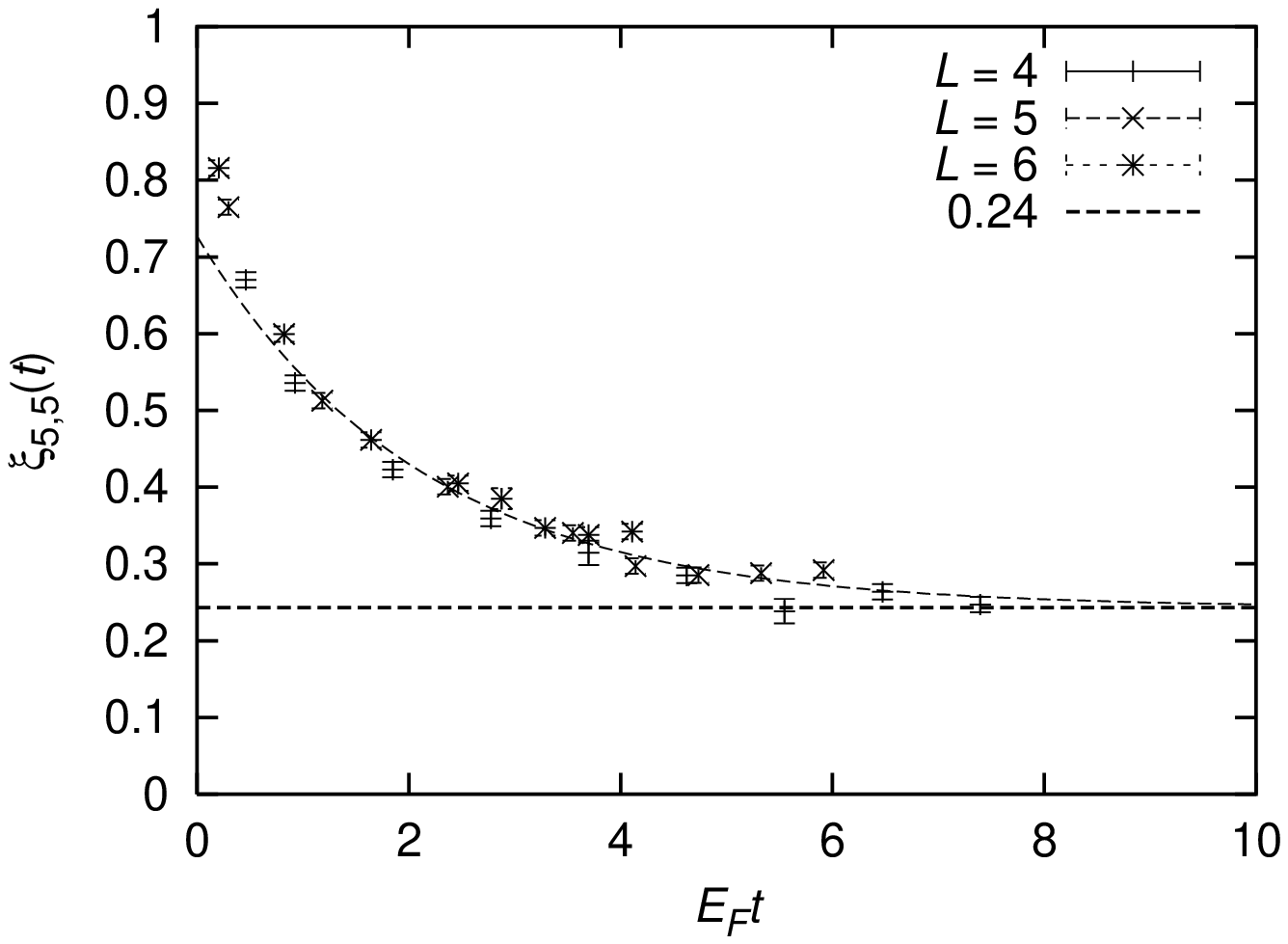}%
\caption{$\xi_{5,5}(t)$ versus $E_{F}t$ for $L=4,5,6$.}%
\label{5+5}%
\end{center}
\end{figure}

\begin{figure}
[ptbptbptb]
\begin{center}
\includegraphics[
height=2.6238in,
width=3.6677in
]%
{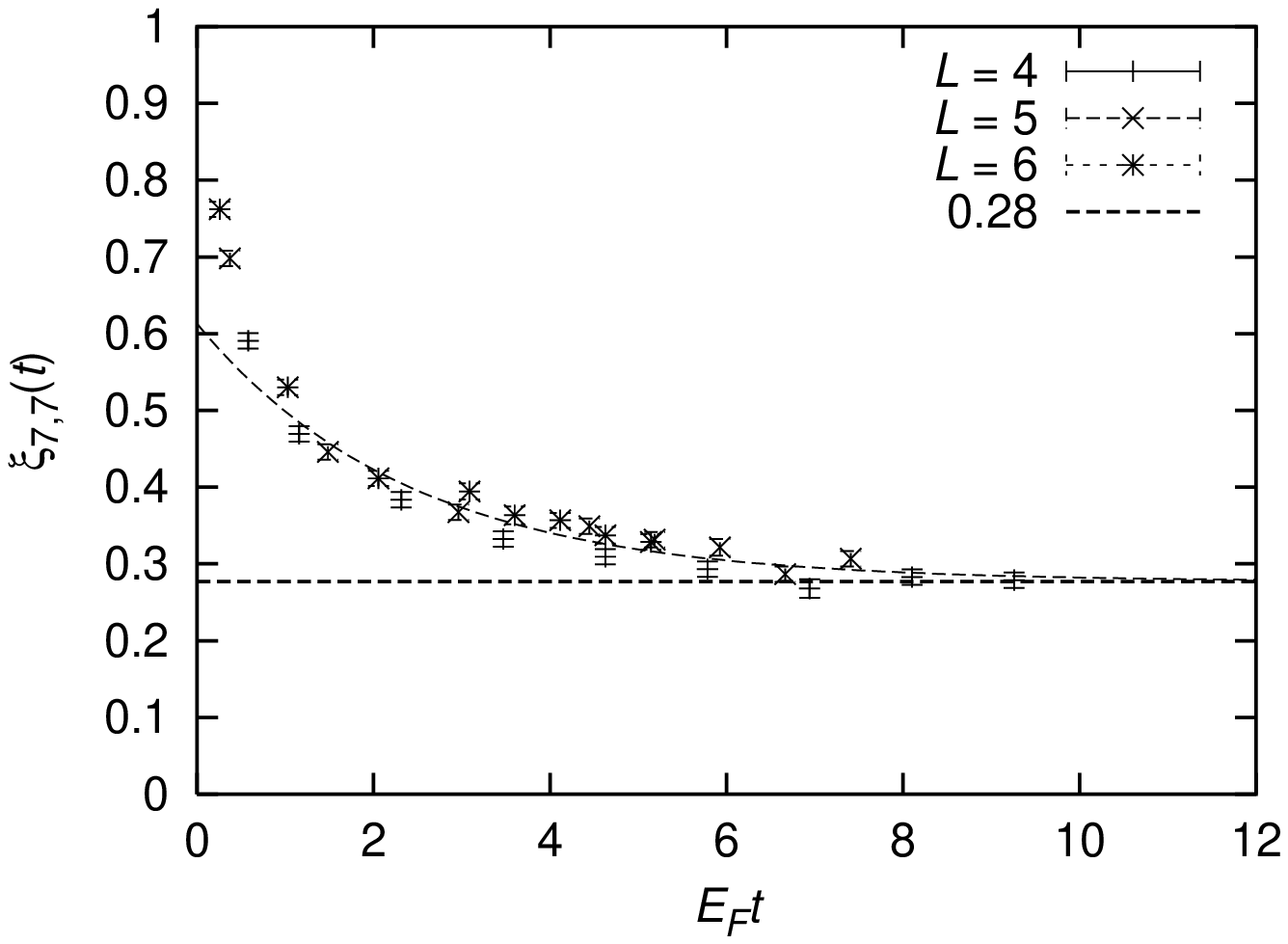}%
\caption{$\xi_{7,7}(t)$ versus $E_{F}t$ for $L=4,5,6$.}%
\label{7+7}%
\end{center}
\end{figure}

\begin{figure}
[ptbptbptbptb]
\begin{center}
\includegraphics[
height=2.6238in,
width=3.6677in
]%
{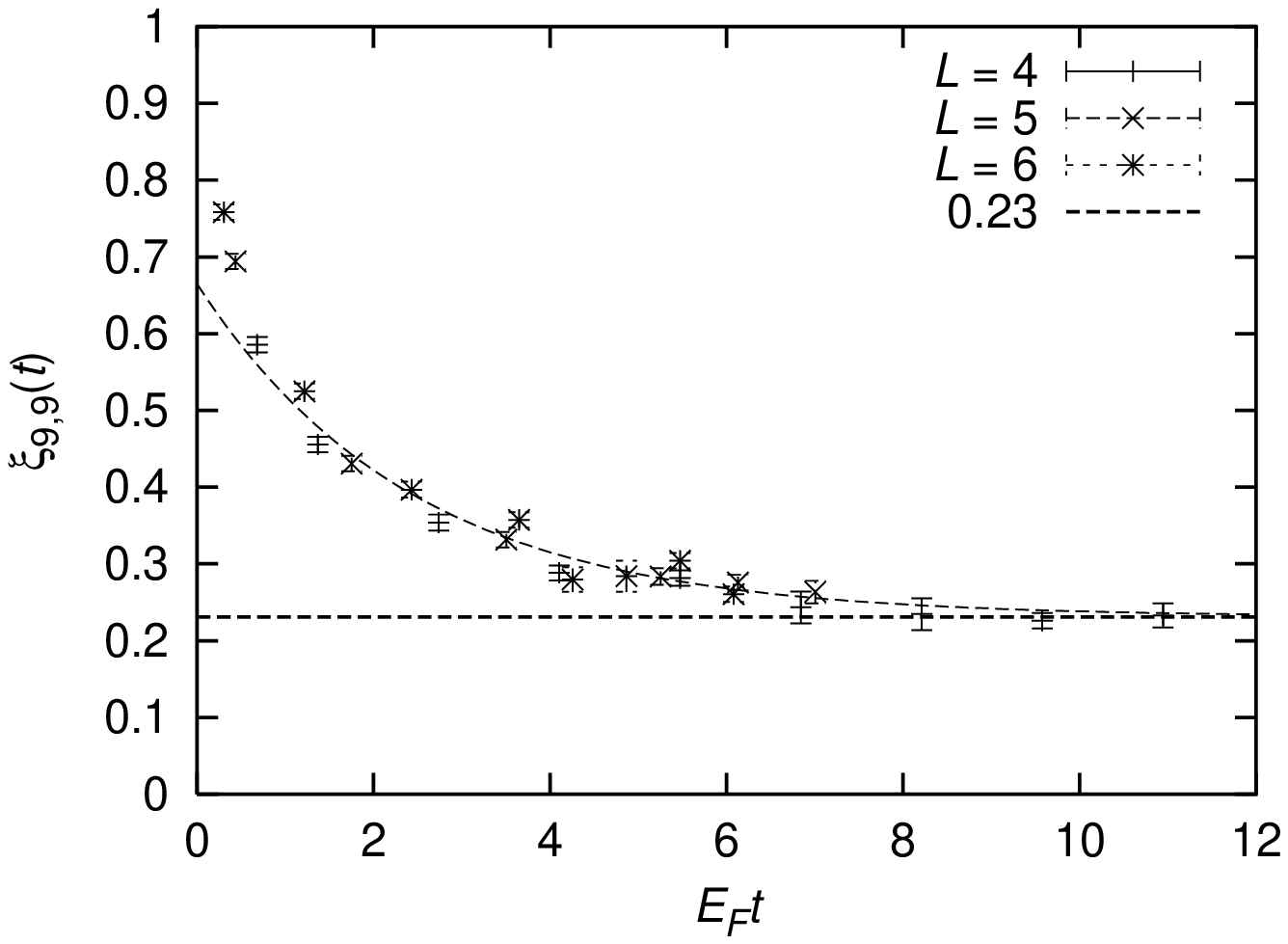}%
\caption{$\xi_{9,9}(t)$ versus $E_{F}t$ for $L=4,5,6$.}%
\label{9+9}%
\end{center}
\end{figure}

\begin{figure}
[ptbptbptbptbptb]
\begin{center}
\includegraphics[
height=2.6238in,
width=3.6677in
]%
{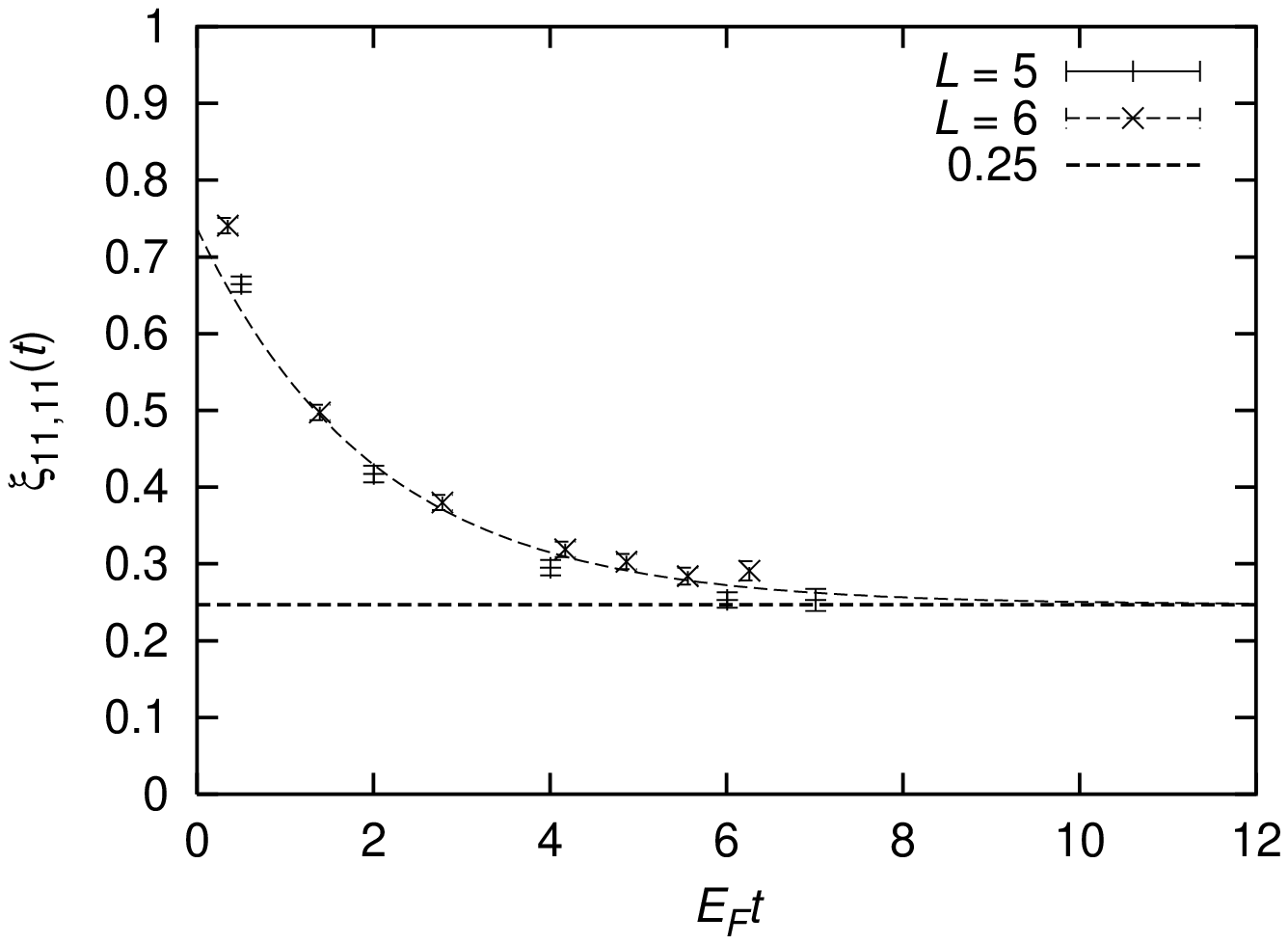}%
\caption{$\xi_{11,11}(t)$ versus $E_{F}t$ for $L=5,6$.}%
\label{11+11}%
\end{center}
\end{figure}

We have performed hybrid Monte Carlo simulations of the $N=3,5,7,9$ systems
with lattice lengths $L=4,5,6$ and the $N=11$ system with lattice lengths
$L=5,6$. \ In Figs. \ref{3+3}, \ref{5+5}, \ref{7+7}, \ref{9+9}, \ref{11+11} we
show $\xi_{N,N}(t)$ as a function of $E_{F}t$ for $N=3$, $5$, $7$, $9$, $11$
respectively.%
In all cases we find agreement in $\xi_{N,N}(t)$ for different values of $L$,
indicating the physics signature of the unitary limit. \ The agreement for
different values of $L$ shows that to a good approximation $L$ is the only
dimensionful scale in the system. The lack of any significant corrections due
to varying $L$ indicates that the lattice discretization error is small and
that the scattering length is essentially infinite. \ This provides a
non-trivial check within the context of the many body system that the physics
is consistent with the unitary limit. \ We find that detuning the coupling
constant away from the unitary limit value by as little as $5\%$ in either
direction produces a measurable disagreement in $\xi_{N,N}(t)$ for different
values of $L$. \ We discuss this further in the next section.

We fit $\xi_{N,N}(t)$ at large $E_{F}t$ to the functional form $b+c\exp
(-\delta\cdot E_{F}t)$ to determine the asymptotic value as $E_{F}%
t\rightarrow\infty$. \ This is the asymptotic form one expects for very large
$E_{F}t$, where $\delta\cdot E_{F}$ is the gap in the energy spectrum above
the ground state energy $E_{N,N}^{0}$.\ \ If there is a non-negligible
coupling to gapless phonon modes, then $\delta$ should go to zero as we
increase the number of particles and volume. \ However it is not clear which
excited states have a non-negligible overlap with our free particle
initial/final state. \ It is also unclear if we are probing at sufficiently
large $E_{F}t$ and with sufficient accuracy to determine $\delta$
reliably.\ \ We hope to resolve these questions with future studies. \ For our
purposes here we consider $\delta$ only as a fit parameter to determine the
asymptotic value for $\xi_{N,N}(t)$ at large $E_{F}t$.

The ratio of the ground state energy of the interacting system to the free
particle ground state energy is given by the large $E_{F}t$ limit of
$\xi_{N,N}(t)$. \ The results for these ratios are shown in Table 2.%
\[%
\genfrac{}{}{0pt}{}{\text{Table 2: \ Results for }E_{N,N}^{0}/E_{N,N}%
^{0,\text{free}}}{%
\begin{tabular}
[c]{|l|l|l|l|l|}\hline
$N=3$ & $N=5$ & $N=7$ & $N=9$ & $N=11$\\\hline
$0.19(2)$ & $0.24(2)$ & $0.28(2)$ & $0.23(2)$ & $0.25(2)$\\\hline
\end{tabular}
}%
\]
From Table 2 we see that the energy ratio for the smallest system, $N=3$, is
somewhat lower than the rest. \ However the ratios for $N\geq5$ are close to a
central value of about $0.25.$ Assuming no large changes to this ratio for
$N>11$, we estimate that $\xi=0.25(3)$.

\begin{figure}
[ptb]
\begin{center}
\includegraphics[
height=3.077in,
width=3.6677in
]%
{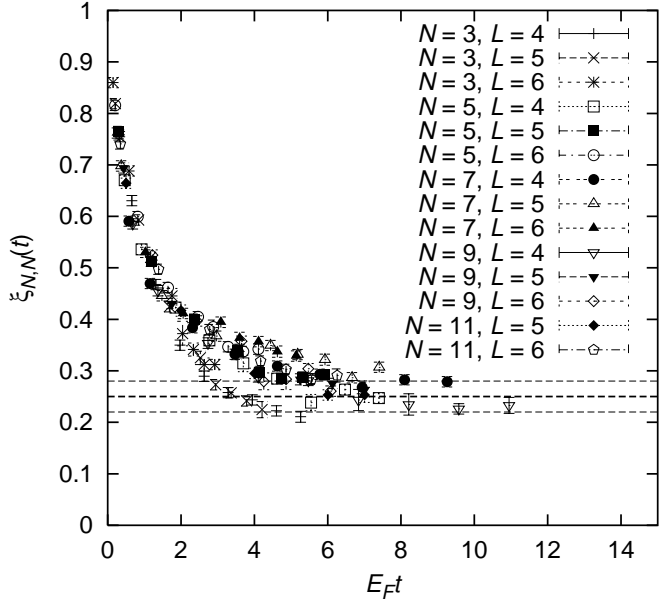}%
\caption{$\xi_{N,N}(t)$ versus $E_{F}t$ for $N=3,5,7,9,11$. \ We have drawn
lines showing the central value and error bounds for the estimate
$\xi=0.25(3)$.}%
\label{all}%
\end{center}
\end{figure}

In Fig. \ref{all} we show simultaneously all of the plots of $\xi_{N,N}(t)$ as
a function of $E_{F}t$.%
We see again that the smallest system, $N=3$, is somewhat different from the
rest, falling off with a faster exponential at large $E_{F}t$. \ The curves
for $N=5,7,9,11$ seem quite similar in both shape and asymptotic value,
suggesting that the curves are already close to the large $N$ limit.

\section{Discussion and additional cross-checks}

Our result $\xi=0.25(3)$ disagrees with the fixed-node Green's function Monte
Carlo results $\xi=0.44(1)$ \cite{Carlson:2003z} and $0.42(1)$
\cite{Astrakharchik:2004}. \ However it is consistent with the requirement
that the fixed-node value sets a variational upper bound on the energy. \ One
explanation of the disagreement could be the difficulty in determining a lower
bound for $\xi$ using Green's function Monte Carlo. \ This requires removing
the nodal constraint and overcoming a sign problem that scales exponentially
with $N$. \ But in any case the disagreement of our final result and published
fixed-node results suggests further tests of the robustness of our method.
\ We consider three additional cross-checks in this section.

The first test we consider is whether or not the non-quadratic lattice
dispersion relation is significantly affecting our results. \ While deviations
were found in nonzero temperature simulations \cite{Lee:2005is}, the fact that
$\xi_{N,N}(t)$ is the same for different values of $L$ suggests that this is
not the case at zero temperature. \ However if we are measuring true continuum
limit behavior, it should be possible to reproduce the same results using a
different lattice action with the same continuum limit. \ To carry out this
cross-check we consider an $O(a^{2})$-improved action with next-to-nearest
neighbor hopping that reproduces the continuum dispersion relation
$p^{2}/(2m)$ up to $O(p^{6})$. \ Instead of the standard action
(\ref{standard}), we use the improved action%
\begin{align}
&  \sum_{\vec{n},i}\left[  c_{i}^{\ast}(\vec{n})c_{i}(\vec{n}+\hat
{0})-e^{\sqrt{-C\alpha_{t}}s(\vec{n})+\frac{C\alpha_{t}}{2}}(1-\frac{15}%
{2}h)c_{i}^{\ast}(\vec{n})c_{i}(\vec{n})\right]  \nonumber\\
&  -\frac{4}{3}h\sum_{\vec{n},l_{s},i}\left[  c_{i}^{\ast}(\vec{n})c_{i}%
(\vec{n}+\hat{l}_{s})+c_{i}^{\ast}(\vec{n})c_{i}(\vec{n}-\hat{l}_{s})\right]
\nonumber\\
&  +\frac{1}{12}h\sum_{\vec{n},l_{s},i}\left[  c_{i}^{\ast}(\vec{n})c_{i}%
(\vec{n}+2\hat{l}_{s})+c_{i}^{\ast}(\vec{n})c_{i}(\vec{n}-2\hat{l}%
_{s})\right]  +\frac{1}{2}\sum_{\vec{n}}\left(  s(\vec{n})\right)  ^{2}.
\end{align}
As before we calculate the interaction coefficient by summing two-particle
scattering bubble diagrams. \ We find in the unitary limit, $C^{\text{phys}%
}=-1.581\times10^{-4}$ MeV$^{-2}.$

\begin{figure}
[ptb]
\begin{center}
\includegraphics[
height=2.6238in,
width=3.6677in
]%
{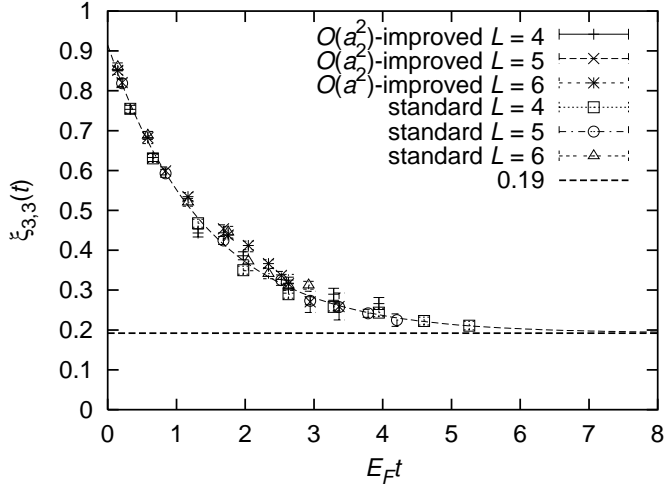}%
\caption{$O(a^{2})$-improved and standard lattice action results for
$\xi_{3,3}(t)$ versus $E_{F}t$ for $L=4,5,6$.}%
\label{improve3+3}%
\end{center}
\end{figure}

The results for $N=3$ are shown in Fig. \ref{improve3+3}. \ We have taken the
standard lattice action results shown in Fig. \ref{3+3} and superimposed the
new $O(a^{2})$-improved lattice data. \ There is almost no difference between
the standard and $O(a^{2})$-improved lattice results. \ This suggests that we
are measuring continuum limit behavior.%

The second test we consider is whether the interaction coefficient can be
retuned to recover the fixed-node value for $\xi$ while approximately
retaining the physics of the unitary limit. \ In order to test this
possibility we consider the $N=5$ system. \ We tune the interaction
coefficient so that for $L=4$ we get $\lim_{t\rightarrow\infty}\xi
_{5,5}(t)=0.42$. \ By trial and error we find that this occurs for
$C^{\text{phys}}=-1.08\times10^{-4}$ MeV$^{-2}$, which is about $90\%$ of the
unitary limit value. \ In the two-particle system this corresponds with a
scattering length of $a_{\text{scatt}}\simeq-7.0$ fm $\simeq-0.035$%
\ MeV$^{-1}.$

\begin{figure}
[ptb]
\begin{center}
\includegraphics[
height=2.6238in,
width=3.6677in
]%
{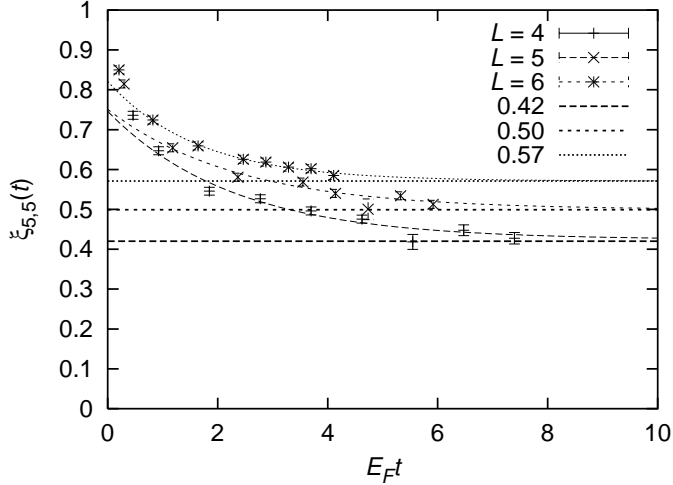}%
\caption{$\xi_{5,5}(t)$ versus $E_{F}t$ for $L=4,5,6$ and $C^{\text{phys}%
}=-1.08\times10^{-4}$ MeV$^{-2}$.}%
\label{detuned5+5}%
\end{center}
\end{figure}

We now simulate the system for $L=5$ and $L=6$. \ The results are shown in
Fig. \ref{detuned5+5}. \ The plots for $\xi_{5,5}(t)$ do not agree for
different $L$, and we find $\lim_{t\rightarrow\infty}\xi_{5,5}(t)=0.42$,
$0.50$, $0.57$ for $L=4,5,6$ respectively. \ From these results it appears
that $\lim_{t\rightarrow\infty}\xi_{5,5}(t)=0.42$ is inconsistent with the
unitary limit.%
In fact we can use this data to estimate the finite scattering length
correction. \ Using $a_{\text{scatt}}\simeq-0.035$\ MeV$^{-1}$ we find%
\begin{equation}
\lim_{t\rightarrow\infty}\xi_{5,5}(t)\approx0.24-0.6\cdot k_{F}^{-1}%
a_{\text{scatt}}^{-1}.
\end{equation}
A more thorough study of the finite scattering length correction to $\xi$ will
be presented in future work.

The third test we consider is whether the same ground state energy can be
extracted using different initial/final states. \ We test this using the $N=7$
system with $L=4$. \ We define $\left\vert \Psi_{7,7}^{2}\right\rangle $ as
the state we get by taking $\left\vert \Psi_{7,7}^{0}\right\rangle $ and
removing the four fermions at momentum $\left(  p_{x},p_{y},p_{z}\right)
=\left(  0,0,\pm\frac{2\pi}{aL}\right)  $ and setting them at higher momentum,
$\left(  p_{x},p_{y},p_{z}\right)  =\left(  \pm\frac{2\pi}{aL},\pm\frac{2\pi
}{aL},0\right)  $. \ We define $\left\vert \Psi_{7,7}^{1}\right\rangle $ in
the same way, except we put only two fermions at the higher momentum while
keeping the total momentum equal to zero.

\begin{figure}
[ptb]
\begin{center}
\includegraphics[
height=2.6238in,
width=3.6677in
]%
{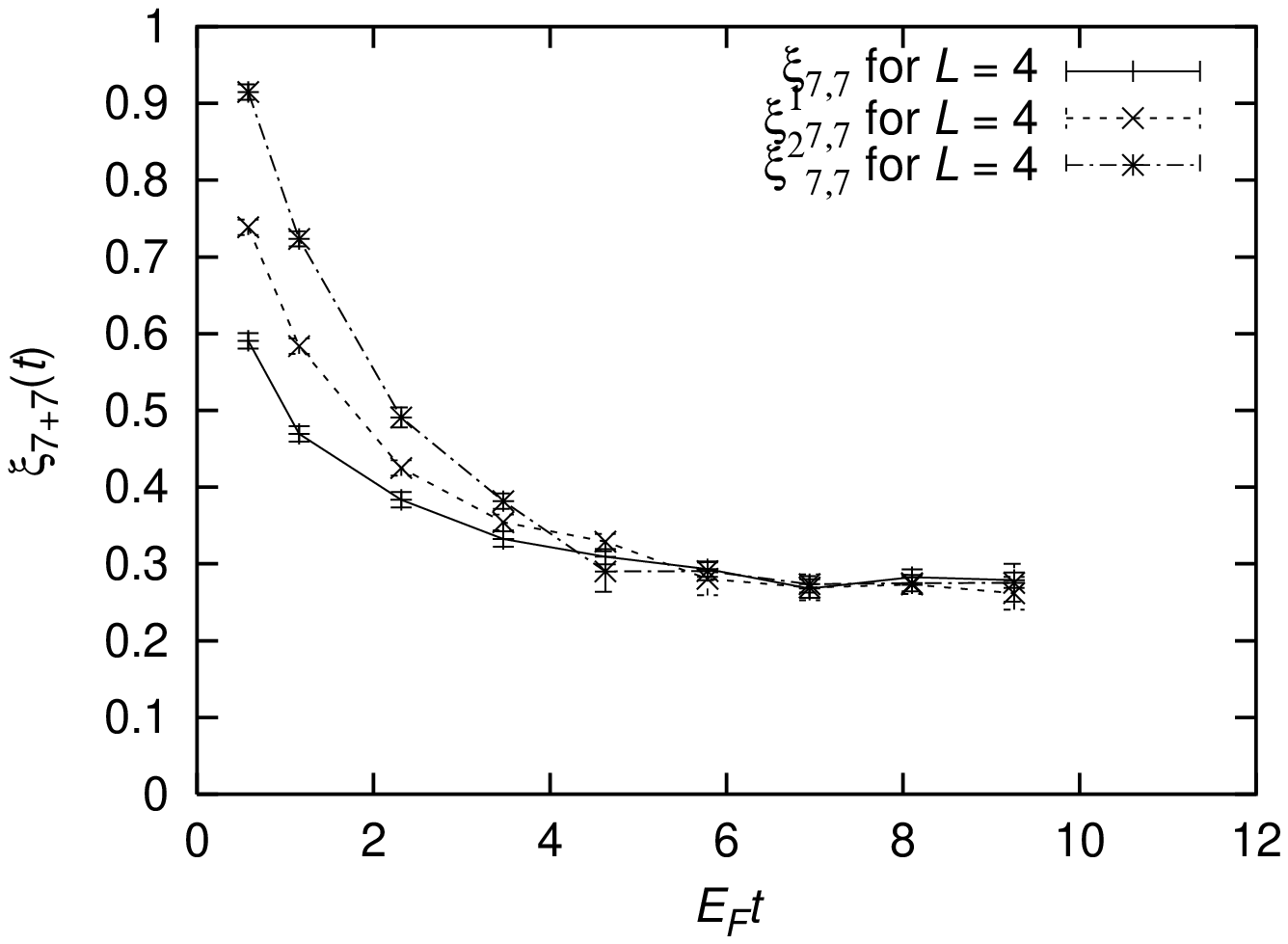}%
\caption{Comparison of $\xi_{7,7}\left(  t\right)  $, $\xi_{7,7}^{1}\left(
t\right)  $, and $\xi_{7,7}^{2}\left(  t\right)  $.}%
\label{7+7excited}%
\end{center}
\end{figure}

We define the following quantities using the excited Slater determinant
states:%
\begin{align}
Z_{7,7}^{1}(t) &  =\left\langle \Psi_{7,7}^{1}\right\vert e^{-Ht}\left\vert
\Psi_{7,7}^{1}\right\rangle ,\quad Z_{7,7}^{2}(t)=\left\langle \Psi_{7,7}%
^{2}\right\vert e^{-Ht}\left\vert \Psi_{7,7}^{2}\right\rangle ,\\
E_{7,7}^{1}(t) &  =-\frac{\partial}{\partial t}\left[  \ln Z_{7,7}%
^{1}(t)\right]  ,\quad E_{7,7}^{2}(t)=-\frac{\partial}{\partial t}\left[  \ln
Z_{7,7}^{2}(t)\right]  ,\\
\xi_{7,7}^{1}(t) &  =\frac{E_{7,7}^{1}(t)}{E_{7,7}^{0,\text{free}}},\quad
\xi_{7,7}^{2}(t)=\frac{E_{7,7}^{2}(t)}{E_{7,7}^{0,\text{free}}}.
\end{align}
The comparison of $\xi_{7,7}(t)$, $\xi_{7,7}^{1}(t),$ and $\xi_{7,7}^{2}(t)$
is shown in Fig. \ref{7+7excited}.%
The convergence of all three results at large $E_{F}t$ suggests that we are
measuring the true ground state energy at large $E_{F}t$.

\section{Summary}

We have measured the ground state energy for $N,N$ spin-1/2 fermions in the
unitary limit in a periodic cube. \ Our results at $N=3,5,7,9,11$ suggest that
for large $N$ the ratio of the ground state energy to that of a free Fermi gas
is $0.25(3)$. \ Our result lies in the middle of the bound $0.07\leq\xi
\leq0.42$ given in \cite{Lee:2005it}.\ \ Although it is inconsistent with the
fixed-node Green's function Monte Carlo results $\xi=0.44(1)$
\cite{Carlson:2003z} and $0.42(1)$ \cite{Astrakharchik:2004}, it is consistent
with the requirement that the fixed-node value sets a variational upper bound
on the energy. \ We have cross-checked the robustness of our results by
comparing with an $O(a^{2})$-improved action, detuning away from the unitary
limit to measure deviations, and using different initial/final states to
reproduce the same ground state energy. \ In this paper we have not addressed
the questions of the existence of superfluidity, long range order, or the
pairing gap. \ However the methods presented here shows some promise at
probing some of these interesting questions.

Acknowledgments: The author is grateful to Thomas Sch\"{a}fer for many helpful
discussions. \ This work is supported by the US Department of Energy grant DE-FG02-04ER41335.

\bibliographystyle{apsrev}
\bibliography{NuclearMatter}

\end{document}